\documentclass[final]{appolb}
\usepackage{float}

\begin{document}
\title{PRESENT STATUS OF REALISTIC SHELL-MODEL CALCULATIONS 
\thanks{Presented at the XXXV Zakopane School of Physics, Zakopane, Poland, September 6-13, 2000 .} }

\author{A. Covello, L. Coraggio, A. Gargano, and N. Itaco
\address{Dipartimento di Scienze Fisiche, Universit\`a di Napoli Federico II \\
and Istituto Nazionale di Fisica Nucleare \\
Complesso Universitario di Monte S. Angelo \\ Via Cintia, 80126 Napoli, Italy} }

\maketitle
\begin{abstract}
The present paper is comprised of two parts. First, we give a brief survey of the theoretical framework for microscopic nuclear structure calculations starting from
a free nucleon-nucleon potential. Then, we present some selected results of a comprehensive study of nuclei near doubly closed shells. In all these shell-model calculations we have made use of realistic effective interactions derived from the Bonn-A nucleon-nucleon potential by means of a $G$-matrix folded-diagram method. We show that this kind of calculation yields a very good agreement with experiment for all the nuclei considered. This leads to the conclusion that realistic effective interactions are now able to describe with quantitative accuracy the properties of complex nuclei.
\end{abstract}

  
\section{Introduction}

A basic input to nuclear shell-model calculations is the model-space effective interaction  $V_{\rm eff}$. Over the past fifty years a variety of empirical effective interactions have been constructed, several of them leading to a very successful description of nuclear structure properties. While this approach has proved to be of great practical value in the interpretation of a large number of experimental data, it is not satisfactory from a first-principle point of view. In fact, a fundamental goal of nuclear structure theory is to understand the properties of complex nuclei in terms of the nucleon-nucleon ($NN$) interaction. This implies the derivation of $V_{\rm eff}$ from the free $NN$ potential.  As is well known, a first step in this direction was taken by Kuo and Brown \cite{kuo66}, who in the mid 1960s derived an $s$-$d$ shell effective interaction from the Hamada-Johnston potential \cite{hamada62}. A few years later, an effective interaction for the lead region was derived by Kuo and Herling \cite{herling72} from the same potential. This was then used in an extensive shell-model study \cite{mcgrory75} of several nuclei around doubly magic $^{208}$Pb. 

Although the results of these early works were more than encouraging, in the following two decades there was much skepticism \cite{elliott88} about the reliability of what had become known as ``realistic" shell-model calculations. As a consequence, this kind of calculation was practically no longer pursued. 

>From the late 1970s on, however, there has been substantial progress towards a microscopic approach to nuclear structure calculations starting from a free $NN$ potential. This has concerned both the two basic ingredients which come into play in this approach, namely the $NN$ potential and the many-body methods for deriving the model-space effective interaction.

As regards the first point, several $NN$ potentials have been constructed, which give a quite good, or even excellent, description of the $NN$ scattering data and are suitable for application in nuclear structure. Regarding the theoretical basis of these potentials,
they can be divided into two categories, according to whether they are based on the meson theory of nuclear forces or are purely phenomenological in nature (except for the one-pion-exchange tail which is included in all potentials). We shall discuss this subject in some detail in subsection 2.1.  

As for the second point, an accurate calculation of the Brueckner $G$-matrix has become feasible while the so-called $\hat Q$-box folded-diagram series for the effective interaction $V_{\rm eff}$ can be summed up to all orders using iterative methods. An outline of the essentials of this derivation will be given in subsection 2.2. 

Based on these improvements, in recent years there has been a revival of interest in shell-model calculations employing effective interactions derived from the free $NN$ potential. In this context, a key issue is how accurate a description of nuclear structure properties can be provided by realistic effective interactions. We have concentrated our efforts during the past five years on this problem, trying to assess systematically the role of these interactions in nuclear structure theory. To this end, we have studied a number of nuclei around doubly magic $^{100}$Sn, $^{132}$Sn, and $^{208}$Pb [6--14]. In particular, we have focused attention on nuclei with few valence particles or holes, since they provide the best testing ground for the basic ingredients of shell-model calculations, especially as regards the matrix elements of the effective $NN$ interaction.
It is our aim here to give a brief survey of the present status of realistic shell-model calculations. While it should be mentioned that calculations similar to ours have been performed for the Sn isotopes and the $N=82$ isotones \cite{holt97,holt98}, the content and conclusions of this paper are based exclusively on our own comprehensive study.

Although we have studied some nuclei with both neutrons and protons in valence shells \cite{andr99}, we shall only consider here nuclei with identical valence nucleons, for which we feel we have now a well defined scenario. In several cases we have also performed calculations using various effective interactions derived from different $NN$ potentials \cite{cov99b}. All the results summarized in this paper, however,  have been obtained by making use of effective interactions derived from the Bonn-A free $NN$ potential. We will come back to this point in the following Section. 

Our presentation is organized as follows. In Sec. 2 we give an outline of the theoretical framework in which our realistic shell-model calculations have been performed. Some selected results are presented in Sec. 3, where we also comment on the whole of our calculations. Sec. 4 presents a summary of our conclusions. 

\section{Theoretical framework}

\subsection{The nucleon-nucleon potential}

As regards the field of $NN$ potentials, we only give here a brief survey of the main aspects relevant to nuclear structure. An historical review and a detailed description of the most recent developments are to be found in Refs. [17--22]. 

>From the theoretical viewpoint, the only quantitative models for the $NN$ interaction are based on meson theory.  Indeed, the early sound theoretical models were the one-boson-exchange (OBE) potentials first developed in the 1960s. Later on, sustained efforts were made to take into account multi-meson exchanges, in particular the $2\pi$-exchange contribution. These efforts have been essentially based on two approaches: dispersion relations and field theory. The work along these two lines, which went on for about a decade, resulted eventually in the Paris potential \cite{lacom80} and in the so called ``Bonn full model" \cite{mach87}, the latter including also contributions beyond $2\pi$. These potentials fit the world $NN$ data below 300 MeV laboratory energy with a $\chi^2$/datum = 3.71 and 1.90, respectively \cite{mach94}. The Bonn full model is, however, an energy-dependent potential, which makes it inconvenient for application in nuclear structure calculations. Therefore, an energy-independent one-boson-parametrization of this potential has been developed \cite{mach89,mach87}, which has become known as ``Bonn A potential". The phase-shift predictions by this potential are very similar to the ones by the Bonn full model with a $\chi^2$/datum of about 2. Bonn-A, like full Bonn, includes nonlocalities leading to a weak tensor force, as measured by the predicted $D$-state probability of the deuteron, $P_D$=4.4\%, to be compared \cite{mach94} with 5.8\% for the Paris potential. 

Over the last ten years or so both the Paris and Bonn-A potentials have been used in nuclear structure calculations. In some cases \cite{andr96,cov97,cov99b,hjorth92,jiang92} comparisons between the results given by these two potentials have been made, which speak in favor of the Bonn-A potential. In particular, from our own calculations for several medium-heavy nuclei \cite{andr96,cov97,cov99b}, it turns out that this potential leads to the best agreement with experiment for all of the nuclei considered.
  
During the last few years, new high-quality $NN$ potentials have been constructed which fit the 1992 Nijmegen database \cite{stoks93} (this contains 1787 $pp$ and 2514 $np$ data below 350 MeV) with a $\chi^2$/datum $\approx 1$. These are the potentials constructed by the Nijmegen group, Nijm-I, NijmII and Reid93 \cite{stoks94}, the Argonne V$_{18}$ potential \cite{wir95} , and the CD-Bonn potential \cite{mach01,mach96}. The first four potentials are purely phenomenological in nature (except for the correct OPE tail) while CD-Bonn is essentially a new version of the OBE Bonn  potential with an additional fit freedom obtained  by adjusting the parameters of the $\sigma$ boson in each partial wave \cite{mach01}.  A comparison between the predictions of these potentials is made in Refs. \cite{machb99,mach00}. In this context, it may be mentioned that in Ref. \cite{mach01} the $\chi^2$/datum produced by these potentials in regard to the 1999 $NN$ database is considered and the conclusion is drawn that the CD-Bonn potential gives the most accurate reproduction of the presently available $NN$ data. 

All the high-precision $NN$ potentials mentioned above use about 45 parameters. The price one has to pay to achieve a very accurate fit of the $NN$ data is well brought out by noting that while Bonn-A has only 13 parameters, which are coupling constants and cutoff masses (see Table A.1 of Ref. \cite{mach89}), CD-Bonn uses 43 free parameters \cite{mach01}. This makes it clear that, to date, high-quality potentials with an excellent $\chi^2$/datum $\approx 1$ can only be obtained within the framework of a substantially phenomenological approach.

Since the high-precision $NN$ potentials fit almost equally well the $NN$ data up to the inelastic threshold, they are essentially identical on-shell. They may largely differ, however, in their off-shell behavior. These off-shell differences may show up in nuclear matter \cite{muth99} and nuclear structure calculations. As mentioned in the Introduction, we are currently studying to which extent nuclear structure calculations depend on the $NN$ potential used as input. From the results obtained so far \cite{cov99b} it turns out that Bonn-A and CD-Bonn give similar results. These two potentials have essentially the same off-shell behavior \cite{mach96}, but show large differences on-shell in terms of the $\chi^2$. The findings of our nuclear structure calculations are quite in agreement with those of Refs. \cite{mach94,mach00} which are based on calculations of the binding energy of the triton. In this connection, it is worth noting that still to date the experimental value (8.48 MeV) of this quantity is best reproduced by the Bonn-A potential, which predicts 8.35 MeV \cite{brand88}. The prediction of the CD-Bonn potential (8.0 MeV) comes rather close to this value while all the other high-precision potentials underbind the triton by about 800 keV. This may be traced to differences in the strength of the tensor force. While Bonn-A and CD-Bonn predict $P_D=4.4$\% and 4.8\%, respectively, the other potentials yield $P_D \approx 5.7$\% \cite{mach00}. These facts suggest that potentials with a weak tensor force may lead to a better description of nuclear structure properties. This is quite an interesting point since differences in $P_D$ may in turn be traced to off-shell differences. We shall not go further into this subject here, but would like to emphasize that it certainly demands a more quantitative understanding.

As already mentioned in the Introduction, the results presented in this paper have all been obtained by using the Bonn-A potential. The brief survey given above should serve mainly the purpose of showing that this potential provides a sound basis not only to test the accuracy of realistic shell-model calculations, but also the meson theory in which it is rooted.

\subsection{The shell-model effective interaction}

In our calculations we make use of two-body effective interactions derived from the  free $NN$ potential  by way of a 
$G$-matrix folded diagram method.  Here, we only outline the essentials of the method and give the references relevant to this short presentation. A detailed description of the derivation of $V_{\rm eff}$ including a more complete list of references can be found in Refs. \cite{kuo95,kuo97}.

As usual, the effective interaction $V_{\rm eff}$ is defined by

\begin{equation}
H_{\rm eff}P \psi_{\mu}= (H_{0}+V_{\rm eff})P \psi_{\mu}=E_{\mu}P \psi_{\mu},
\end{equation}

\noindent
where the eigenvalues $E_{\mu}$ are a subset of the eigenvalues of the 
original Hamiltonian in the full space, $\mu=1,2 \ldots d$ with $d$ denoting the 
dimension of the model space. $H_{0}=T+U$ is the unperturbed Hamiltonian, $T$ being the kinetic energy and $U$ an auxiliary
potential introduced to define a convenient single-particle basis. $U$ is
usually chosen to be the harmonic oscillator potential. Here $P$ 
denotes the projection
operator onto the chosen model space, which generally consists of a major
shell above the doubly closed core.

The first problem one is confronted with
in the derivation of $V_{\rm eff}$ is  that all modern $NN$
potentials have a strong repulsive core, which makes perturbation calculations
meaningless. This difficulty can be overcome by introducing 
the $G$ matrix defined by the integral equation \cite{krenc76}

\begin{equation}
G(\omega)=V+VQ_2 \frac{1}{\omega-Q_2TQ_2}Q_2G(\omega),
\end{equation}

\noindent
where $V$ represents the $NN$ potential, $T$ is the two-nucleon kinetic
energy, and $\omega$, commonly referred to as starting-energy, is  
the unperturbed energy of the interacting nucleons.
The operator $Q_{2}$ is the Pauli exclusion operator for two interacting
nucleons, to make sure that the intermediate states of $G$ must not only
be above the filled Fermi sea but also outside the model space within
which equation (2) is to be solved. Thus the Pauli operator
$Q_{2}$ is dependent on the model space chosen, and so is the
corresponding $G$ matrix. Note that the operator $Q_{2}$ is defined in terms
of harmonic oscillator wave functions while plane-wave functions are
employed for the intermediate states of the $G$ matrix. As regards the
harmonic oscillator parameter $\hbar \omega$ we have used the values given by the
formula
$\hbar \omega= (45 A^{-1/3} -25 A^{-2/3}$) MeV, where $A$ is the mass
number of the core.

The operator $Q_{2}$ is specified, as discussed in \cite{krenc76},
by three numbers $n_{1}$,
$n_{2}$, and $n_{3}$, each representing a single-particle orbital
(the orbits are numbered starting from the bottom of the oscillator well).
In particular, $n_{1}$ is the number of orbitals below the
Fermi surface of the doubly magic core, $n_{2}$  fixes the orbital
above which  the passive single-particle states start, and $n_{3}$ denotes 
the number 
of orbitals of the full space. In principle, this last number should 
be infinite. 
In practice, as we shall see in the following, it is 
chosen to be a large but finite number. 

It should be noted that in the calculation of $G$ the space of active 
single-particle states, 
which is defined as the number of levels
between $n_{1}$ and $n_{2}$, may be different from the model space within 
which $V_{\rm eff}$ is defined. Several arguments for choosing 
the former larger
than the latter are given in \cite{krenc76}. Generally, $n_{2}$ is fixed so as
to include two major shells above the Fermi surface.  
In our calculation for the lead region, however, we have found that 
a substantially better 
agreement with experiment is obtained when $n_{2}$ is increased 
from two to three shells above the $n_{1}-$th orbit \cite{cor99}.   

An accurate calculation 
of the plane-wave $G$ matrix (2) is feasible by using the Tsai-Kuo 
method \cite{tsai72}. This method gives the exact 
solution of $G$ as

\begin{equation}
G=G_{F}+\Delta G.
\end{equation}

\noindent
The free $G$ matrix is  
\begin{equation}
G_{F}=V+V\frac{1}{\omega -T}G_{F},
\end{equation}

\noindent
and the Pauli correction term $\Delta G$ is given by

\begin{equation}
\Delta G=-G_{F}\frac{1}{e}P_2\frac{1}{P_2(\frac{1}{e}+
\frac{1}{e}G_{F}\frac{1}{e})P_2}P_2\frac{1}{e}G_{F},
\end{equation}

\noindent
with $e\equiv (\omega - T)$ and $P_2 = 1 - Q_2$.
    
The $G_{F}$ matrix does not contain the Pauli exclusion operator and
hence its calculation is relatively convenient. To compute $\Delta G$,
we have to perform some matrix operations within the 
model space $P_{2}$. The only necessary  approximation in the Tsai-Kuo
method is to make $n_{3}$ a finite number. It was shown \cite{krenc76}, 
however, that
a sufficiently large value of $n_{3}$ ensures the accuracy of this 
approximation. 

As a last point, it is worth noting
that for asymmetric doubly closed cores the $Q_{2}$ operators for
protons and neutrons are different, as the valence-proton and -neutron orbits
outside the core are different. In this case, the calculation of $G$ is more
complicated \cite{cov97}. 
In fact, one has first to calculate $G_{F}$ in a proton-neutron
representation. Then $\Delta G$, which depends on the $P_{2}$ operator, has
to be calculated for protons and neutrons separately.

Using the above $G$ matrix we can now calculate  $V_{\rm
eff}$ in the model space.
This  interaction, which is energy independent,  can be written 
schematically in operator form as \cite{kuo80}

\begin{equation}
V_{\rm eff} = \hat{Q} - \hat{Q'} \int \hat{Q} + \hat{Q'} \int \hat{Q} \int
\hat{Q} - \hat{Q'} \int \hat{Q} \int \hat{Q} \int \hat{Q} + ~...~~,
\end{equation}

\noindent
where $\hat{Q}$ (referred to as $\hat{Q}$-box) is a vertex function composed 
of irreducible  linked diagrams in $G$, and the integral sign represents a
generalized folding operation .
$\hat{Q'}$ is obtained from $\hat{Q}$ by removing terms of first order in the
reaction matrix $G$. In our calculations we take the $\hat{Q}$-box to 
be composed of $G$-diagrams through second order. They are precisely the
 seven first- and second-order diagrams considered by
Shurpin {\em et al.} \cite{shurp83}.
Note that for nuclei with valence holes the calculation of the 
$\hat{Q}$-box  is somewhat different. In fact, in this case
the external valence-particle lines appearing in the diagrams of \cite{shurp83} have to 
be replaced with valence-hole lines \cite{cor00}.

After the $\hat{Q}$-box is calculated, $V_{\rm eff}$ is
obtained by summing up the folded-diagram series (6) to all orders by 
means of the Lee-Suzuki iteration method \cite{suzuki80}.
This last step can be performed in an essentially exact way for a given
$\hat{Q}$-box.

It is worth mentioning that the $V_{\rm eff}$ obtained with the procedure  
described above represents the effective interaction only between 
two-valence particles or holes. This effective interaction contains
one- and two-body terms. It is customary, however, to use a subtraction 
procedure \cite{shurp83} so that only the two-body terms are retained. 
As far as the  one-body terms are concerned,
it is assumed that they modify the unperturbed part of
the Hamiltonian, which justifies the use of experimental single-particle 
energies.

As regards the electromagnetic observables, we have calculated them
by making use of effective operators \cite{mav66,krenc75} which 
take into account core-polarization effects.
More precisely, by using a diagrammatic description as in Ref. \cite{mav66},
we have only included first-order diagrams in $G$. 
This implies that folded-diagram renormalizations are not necessary
\cite{krenc75}. 

\begin{table}[H]
\begin{flushright}
TABLE I
\end{flushright}
\vskip -.4cm
\noindent
{\small Experimental and calculated low-energy levels in $^{130}$Sn.}                   
\begin{center}
\begin{tabular}{ccc|ccccc}
\hline
 & & & & \multicolumn{3}{c} {$E$ (MeV)} &  \\
 & $J^{\pi}$ & & & & & & \\ 
 & & & & Exp. & & Calc. & \\
\hline
 & & & & & & & \\
& $0^+$ & & & 0.000 & & 0.000 & \\
& $2^+$ & & & 1.221 & & 1.353 & \\
& $7^-$ & & & 1.947 & & 1.814 & \\
& $4^+$ & & & 1.996 & & 2.063 & \\
& $5^-$ & & & 2.085 & & 1.937 & \\
& $4^-$ & & & 2.215 & & 1.990 & \\
& $6^+$ & & & 2.257 & & 2.244 & \\
& $8^+$ & & & 2.338 & & 2.324 & \\
& $10^+$ & & & 2.435 & & 2.402 & \\
\end{tabular}
\end{center}
\end{table}

\vskip -1cm
\begin{table}[H]
\begin{flushright}
TABLE II
\end{flushright}
\vskip -.4cm
\noindent
{\small Experimental and calculated low-energy levels in $^{134}$Te.}                   
\begin{center}
\begin{tabular}{ccc|ccccc}
\hline
 & & & & \multicolumn{3}{c} {$E$ (MeV)} &  \\
 & $J^{\pi}$ & & & & & & \\ 
 & & & & Exp. & & Calc. & \\
\hline
 & & & & & & & \\
& $0^+$ & & & 0.000 & & 0.000 & \\
& $2^+$ & & & 1.279 & & 1.299 & \\
& $4^+$ & & & 1.576 & & 1.583 & \\
& $6^+$ & & & 1.691 & & 1.721 & \\
& $6^+$ & & & 2.398 & & 2.371 & \\
& $2^+$ & & & 2.462 & & 2.553 & \\
& $4^+$ & & & 2.554 & & 2.584 & \\
& $1^+$ & & & 2.631 & & 2.515 & \\
& $5^+$ & & & 2.727 & & 2.671 & 
\end{tabular}
\end{center}
\end{table}

\section{Review of selected results}

To illustrate the quantitative accuracy of realistic shell-model calculations
in describing the spectroscopic properties of nuclei near closed shells,
we report here  some results of the study we have performed over
the past few years. These are to be viewed as samples of the whole body of
results we have obtained for medium- and heavy-mass nuclei having or
lacking few identical particles with respect to double shell closures.
\begin{table}[H]
\begin{flushright}
TABLE III
\end{flushright}
\vskip -.4cm
\noindent
{\small Experimental and calculated low-energy levels in $^{206}$Pb.}                   \begin{center}
\begin{tabular}{crc|ccccc}
\hline
 & & & & \multicolumn{3}{c} {$E$ (MeV)} &  \\
 & $J^{\pi}$ & & & & & & \\ 
 & & & & Exp. & & Calc. & \\
\hline
 & & & & & & & \\
& $0^+$ & & & 0.000 & & 0.000 & \\
& $2^+$ & & & 0.803 & & 0.804 & \\
& $0^+$ & & & 1.165 & & 1.036 & \\
& $3^+$ & & & 1.340 & & 1.259 & \\
& $2^+$ & & & 1.467 & & 1.273 & \\
& $4^+$ & & & 1.684 & & 1.693 & \\
& $1^+$ & & & 1.704 & & 1.546 & \\
& $2^+$ & & & 1.784 & & 1.697 & \\
& $4^+$ & & & 1.998 & & 1.934 & \\
& $2^+$ & & & 2.148 & & 2.124 & \\
& $^a 3^+$& & & 2.197 & & 2.142 & \\
& $7^-$ & & & 2.200 & & 2.156 & \\
& $^a 1^+$ & & & 2.236 & & 2.101 & \\
& $0^+$ & & & 2.315 & & 2.135 & \\
& $6^-$ & & & 2.384 & & 2.312 & \\
& $2^+$ & & & 2.423 & & 2.280 & \\
\end{tabular}
\end{center}
{\small $^a$ This spin-parity assignment is suggested by our calculations (see Ref. \cite{cor98}).}
\end{table}
As already pointed out in the preceding Sections,
all the results presented here have been obtained by making use
of effective interactions derived from the Bonn-A free $NN$ potential.
Regarding the single-particle or {\mbox -hole} energies, we have taken them from
the experimental spectra of the corresponding single-particle or -hole valence
nuclei, wherever available. This is indeed the case of the
four nuclei $^{130}$Sn, $^{134}$Te, $^{206}$Pb, and $^{210}$Po, for which
we report here  the calculated spectra.  Actually, 
only for the $N=50$ isotones and the light Sn isotopes we had to
use a different procedure to fix the single-hole
and single-particle spectrum, respectively.
As regards the electromagnetic 
observables, no use has been made of empirical effective charges since, as 
mentioned in subsection 2.1, the effective operators needed for their calculation 
have been derived in a microscopic way.

In Tables I-IV we compare the experimental \cite{nndc,omtvedt95} and 
calculated low-energy spectra
of  $^{130}$Sn, \ $^{134}$Te, \ $^{206}$Pb, and  $^{210}$Po. An example of the kind 
\begin{table}[H]
\begin{flushright}
TABLE IV
\end{flushright}
\vskip -.4cm
\noindent
{\small Experimental and calculated low-energy levels in $^{210}$Po.}                   
\begin{center}
\begin{tabular}{ccc|ccccc}
\hline
 & & & & \multicolumn{3}{c} {$E$ (MeV)} &  \\
 & $J^{\pi}$ & & & & & & \\ 
 & & & & Exp. & & Calc. & \\
\hline
 & & & & & & & \\
& $0^+$ & & & 0.000  & & 0.000 & \\
& $2^+$ & & & 1.181 & & 1.130 & \\
& $4^+$ & & & 1.427 & & 1.395 & \\
& $6^+$ & & & 1.473 & & 1.493 & \\
& $8^+$ & & & 1.557 & & 1.555 & \\
& $8^+$ & & & 2.188 & & 2.179 & \\
& $2^+$ & & & 2.290 & & 2.292 & \\
& $6^+$ & & & 2.326 & & 2.367 & \\
& $4^+$ & & & 2.382 & & 2.394 & \\
& $1^+$ & & & 2.394 & & 2.220 & \\
& $5^+$ & & & 2.403 & & 2.422 & \\
& $3^+$ & & & 2.414 & & 2.380 & \\
& $7^+$ & & & 2.438 & & 2.437 & 
\end{tabular}
\end{center}
\end{table}
\vskip -1cm

\begin{table}[H]
\begin{flushright}
TABLE V
\end{flushright}
\vskip -.4cm
\noindent
{\small Experimental and calculated $B(E2)$ values (W.u.), {\it Q} moments 
($e {\rm b}$), and $\mu$ moments (nm) in $^{210}$Po.}
\begin{center}
\begin{tabular}{c|ccc}
\hline
 & & & \\
 Quantity & Exp.  & &  Calc. \\
 & & & \\
\hline 
 & & & \\
$B(E2;2_1^+ \rightarrow 0_1^+)$ &  $0.56 \pm 0.12 $ & & 3.55 \\ 
$B(E2;4_1^+ \rightarrow 2_1^+)$ &  $4.53 \pm 0.15 $ & & 4.46 \\ 
$B(E2;6_1^+ \rightarrow 4_1^+)$ &  $3.00 \pm 0.12 $ & & 3.07 \\ 
$B(E2;8_1^+ \rightarrow 6_1^+)$ &  $1.10 \pm 0.05 $ & & 1.25 \\ 
$\mu (6_1^+ )$ & $\pm 5.48 \pm 0.05$ & & +5.29 \\
$\mu (8_1^+ )$ & $ + 7.35 \pm 0.05$ & & +7.06 \\
$\mu (11_1^- )$ & $ + 12.20 \pm 0.09$ & & +13.12 \\
$Q (8_1^+ )$ & $ - 0.552 \pm 0.020 $ & & $- 0.588$ \\
$Q (11_1^- )$ & $ - 0.86 \pm 0.11$ & & $-0.92$ 
\end{tabular}
\end{center}
\end{table}
\noindent of agreement between theory and experiment \cite{nndc,becker91}
for the electromagnetic properties  is given in Table V, 
where the moments and the $E2$ transition rates  for $^{210}$Po are reported. 
We see that the observed excitation energies in all the four nuclei are very well reproduced by our calculations, the discrepancy being less than 
100 keV for most of the states. The agreement between experiment and theory
is also very good for the electromagnetic properties of $^{210}$Po, with the
only exception of $B(E2; 2^{+}_{1} \rightarrow 0^{+}_{1})$. 
It should be mentioned, however, that our value is consistent with that of a 
previous calculation \cite{zwarts85}, 
while the experimental one  has been derived  in a way \cite{ellegaard73}
which makes it rather unreliable \cite{cor99}.  
\begin{table}[H]
\begin{flushright}
TABLE VI
\end{flushright}
\vskip -.4cm
\noindent
{\small Root mean square deviation $\sigma$. See text for details.}
\begin{center}
\begin{tabular}{clc|ccc}
\hline
 & & & & &\\
 & Nucleus & & Number of levels &  $\sigma$(keV) & \\
 & & & & & \\
\hline 
 & & & & & \\
 & $^{96}$Pd & & 7               & 122 &   \\
 & $^{97}$Ag & & 4               & 108  & \\
 & $^{98}$Cd & & 4               & 107 & \\
 & $^{102}$Sn & & 3               & 92  & \\
 & $^{104}$Sn & & 5               & 165 & \\
 & $^{130}$Sn & & 8               & 119 & \\
 & $^{134}$Te & & 15               & 127 &  \\
 & $^{135}$I & & 10               & 59 &  \\
 & $^{136}$Xe & & 14              & 90 & \\
 & $^{204}$Pb & & 32              & 122 & \\
 & $^{205}$Pb & & 17              & 64 & \\
 & $^{206}$Pb & & 24              & 94 & \\
 & $^{210}$Po & & 37               & 87 &  \\
 & $^{211}$At & & 19              & 64  & \\
 & $^{212}$Rn & & 13              & 85 &  \\
\end{tabular}
\end{center}
\end{table}
More complete spectra of the four nuclei considered here may be found 
in Refs. \cite{andr97,cor98,cov99,cov99b,cor99}, while  the results obtained for 
all other nuclei included in our study  are given  in Refs. \cite{andr96,andr97,cov97,cor98,cov99b,cor99,cor00}. 
We also refer the reader to these papers for  a detailed analysis of the 
spectroscopic properties of the various nuclei. 
It seems appropriate, however,  to show here the
overall agreement between our
results and the experimental data, at least as regards the energy levels. 
A quantity which provides  such an information is the  
{\it rms} deviation  $\sigma$ \cite{sigma}. In Table VI we report
the $\sigma$ values for all the nuclei we have studied with at 
most four identical 
valence particles or holes in the region of doubly magic $^{100}$Sn, 
$^{132}$Sn, and $^{208}$Pb.
The number of levels included in the calculation of $\sigma$ is also 
reported. 
It should be noted that the very limited number of data for the
lighter nuclei reflects the experimental
situation. In fact, these nuclei lie well away from the valley
of stability and only recently some experimental information on their
spectroscopic properties has become available.
In any case, we see that the $\sigma$ values are all very small,
7 out of 15 being less than 100 keV and only one  reaching  165 keV.
\begin{table}[H]
\begin{flushright}
TABLE VII
\end{flushright}
\vskip -.4cm
\noindent
{\small Experimental and calculated ground-state binding energies (MeV).}
See text for comments.
\begin{center}
\begin{tabular}[H]{cc|ccccc}
\hline
 &  & & \multicolumn{3}{c} {$B$} &  \\
 & Nucleus & &  & & & \\
 & & & Exp. & & Calc. & \\
\hline
 & & & & & & \\
& $^{98}$Cd & & $-3.98 \pm 0.48 $  & & $-4.56 \pm 1.33 $ & \\
& $^{102}$Sn & & $24.03 \pm 0.59 $  & & 23.60 $\pm 1.33$ & \\
& $^{130}$Sn & & $-12.52 \pm 0.04 $  & & $-12.82 \pm 0.06$ & \\
& $^{134}$Te & & $20.56 \pm 0.04 $  & & 20.62 $\pm 0.07 $ & \\
& $^{206}$Pb & & $-14.12 \pm 0.00 $  & & $-14.12 \pm 0.01 $ & \\
& $^{210}$Po & & $8.78 \pm 0.00 $  & & 8.79 $\pm 0.01 $& 
\end{tabular}
\end{center}
\end{table}
Finally, for $^{98}$Cd, $^{102}$Sn, $^{130}$Sn, $^{134}$Te, $^{206}$Pb,
and $^{210}$Po we show in Table VII the comparison between the
observed \cite{audi93,fogelberg99} and  calculated ground-state 
binding energies relative to 
the closest doubly closed core. For the absolute scaling of the six sets of
single-particle or single-hole energies, the
mass excess values for
nuclei with one particle or one hole with respect to $^{100}$Sn, $^{132}$Sn,
and $^{208}$Pb are needed. These have been taken from Refs. \cite{audi93,fogelberg99}. As regards the Coulomb
interaction between the valence protons in $^{98}$Cd, $^{134}$Te, and
$^{210}$Po, we assume that its contribution is equal to the matrix element 
of the Coulomb force between the two-proton states $(g_{9/2})^{2}_{J=0^+}$,
 $(g_{7/2})^{2}_{J=0^+}$, and   $(h_{9/2})^{2}_{J=0^+}$, respectively.
>From Table VII we see that an excellent agreement with experiment is obtained,
the calculated values falling within the error bars for all the nuclei 
considered.

\section{Summary and outlook}

In this paper, we have tried to give a self-contained review of the present status of realistic shell-model calculations, as it appears to us through the work we have carried out in this field during the past five years. The main aim of this work has been to give an answer to the crucial question of whether this kind of calculation is able to provide a quantitative description of nuclear structure properties. We think we can now definitely answer this question in the affirmative with a word of caution. As already mentioned  in the Introduction, the main body of our calculations has until now concerned nuclei with identical valence nucleons. As a consequence, we may only claim to have a stringent test of the isospin $T=1$ matrix elements of the effective interaction. A careful test of the $T=0$ matrix elements is of course equally important. We should point out, however, that in a recent study \cite{andr99} of the doubly odd nucleus $^{132}$Sb we have obtained results which are as good as those regarding like nucleon systems. Along the same lines we are currently studying other nuclei with both neutrons and protons outside closed shells.

Another main question relevant to microscopic nuclear structure calculations is the extent  to which they depend on the $NN$ potential used as input. As briefly discussed in subsection 2.1, we are also trying to explore this problem. We only emphasize here that
the Bonn-A potential, which we have used as initial input in all the calculations reported in Sec. 3, yields results which are similar to (in some cases even somewhat better than)  those produced by its modern, high-precision, version CD-Bonn.  
This indicates that, as far as nuclear structure is concerned, the $\chi^2$ produced by an $NN$ potential is not, within reasonable limits, the most important aspect. In other words, on-shell differences have little influence on nuclear structure results. 

To conclude, we may say that the stage for realistic shell-model calculations is by now well set, that is to say the way is finally open to a more fundamental approach to the nuclear shell model than the traditional, empirical one. From a first-principle point of view, however, we should be aware that a substantial theoretical progress in the field of $NN$ interaction is still in demand. This may not be necessarily achieved within the framework of the traditional meson theory. Suffice here to mention a promising alternative approach, the chiral effective theory, which is actively being pursued \cite{epel00}. 

\vskip 10pt

The results presented in this paper are part of a research project carried out in collaboration with T.T.S. Kuo. This work was supported in part by the Italian Ministero dell'Universit\`a e della Ricerca Scientifica e Tecnologica (MURST). N.I. would like to thank the European Social Fund for financial support.


\begin{thebibliography}{99}
\bibitem{sigma} We define 
$\sigma=\{(1/N_{d})\sum_{i}[E_{exp}(i)-E_{calc}(i)]^{2}\}^{1/2}$,
where $N_{d}$ is the number of data.
\end{thebibliography}
\end{document}